\documentclass{article}

\usepackage{amsmath}
\usepackage{amssymb}

\usepackage{graphicx}
\usepackage{siunitx}
\usepackage{colortbl}
\usepackage{xcolor}
\usepackage{booktabs}
\usepackage{multirow}
\usepackage{array}
\usepackage{soul}

\usepackage[accepted]{icml2020}

\icmltitlerunning{Conditional Normalizing Flows for Low-Dose Computed Tomography Image Reconstruction}

\begin{document}

\bibliographystyle{icml2020}

\twocolumn[
\icmltitle{Conditional Normalizing Flows for \\ Low-Dose Computed Tomography Image Reconstruction}
\icmlsetsymbol{equal}{*}

\begin{icmlauthorlist}
\icmlauthor{Alexander Denker}{to}
\icmlauthor{Maximilian Schmidt}{to}
\icmlauthor{Johannes Leuschner}{to}
\icmlauthor{Peter Maass}{to}
\icmlauthor{Jens Behrmann}{to}
\end{icmlauthorlist}

\icmlcorrespondingauthor{Alexander Denker}{adenker@uni-bremen.de}
\icmlcorrespondingauthor{Maximilian Schmidt}{maximilian.schmidt@uni-bremen.de}
\icmlcorrespondingauthor{Johannes Leuschner}{jleuschn@uni-bremen.de}

\icmlaffiliation{to}{University of Bremen, Center for Industrial Mathematics}
\icmlkeywords{Machine Learning, ICML}

\vskip 0.3in

]

\printAffiliationsAndNotice{}

\begin{abstract}
    {
    Image reconstruction from computed tomography (CT) measurement is a challenging statistical inverse problem since a high-dimensional conditional distribution needs to be estimated. Based on training data obtained from high-quality reconstructions, we aim to learn a conditional density of images from noisy low-dose CT measurements. To tackle this problem, we propose a hybrid conditional normalizing flow, which integrates the physical model by using the filtered back-projection as conditioner. We evaluate our approach on a low-dose CT benchmark and demonstrate superior performance in terms of structural similarity of our flow-based method compared to other deep learning based approaches.
    }
\end{abstract}

\section{Introduction}
\label{sec:introduction}
Many important applications in medical imaging, such as computed tomography (CT) or magnetic resonance imaging (MRI), can be formulated as an inverse problem. The inverse problem consists in the reconstruction of an internal image of a patient based on radiological data. Many of these applications are ill-posed inverse problems, as small measurement errors can result in large errors in the reconstruction. In a classical way, an inverse problem is often formulated as follows: A forward operator $A: X \rightarrow Y$ maps the image $x^\dagger$ to (noisy) measurements
\begin{align}
    y^\delta = Ax^\dagger + \mu,
    \label{eq:inverse_problem}
\end{align}
where $\mu$ describes the noise. The research in inverse problems is focused in particular on developing algorithms for obtaining stable approximations of the true solution $x^\dagger$. In order to cover the uncertainties that occur especially with ill-posed problems, the theory of Bayesian inversion considers the posterior distribution $p(x|y^\delta)$ \citep{Dashti.2272013}. This posterior is the conditional density of the image $x$ conditioned on the measurements $y^\delta$. 

The main task in statistical inverse problems is to approximate this high-dimensional conditional distribution. For high-dimensional, structured images, like they arise in CT, this is a challenging process. In the field of density estimation, conditional normalizing flows (NF) \citep{Winkler.11292019, Ardizzone.742019} allow to learn expressive conditional densities by maximum likelihood training. Since the physical model is known in CT (Eq.~\ref{eq:radon_trafo}), we propose a hybrid approach which integrates model-based reconstruction with conditional NFs.

\begin{figure}
\centering
\includegraphics[width=\columnwidth]{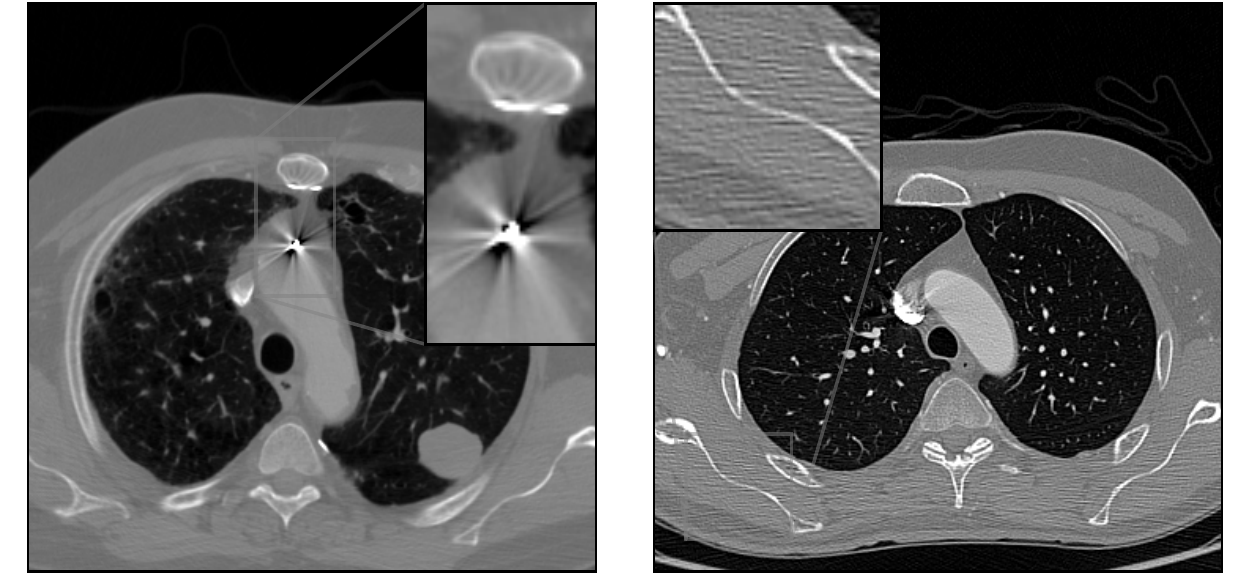}
\caption[ArtifactSamples]{Ground truth samples from the LoDoPaB-CT dataset containing artifacts. These errors stem from the reconstruction technique that was used on the normal-dose measurements.}
	\label{fig:ArtifactSamples}
\end{figure}

In many CT image reconstruction tasks the mean squared error (MSE) is used \citep{chen2017convnet_ct, he2020iradon}, which, however, has many known limitations \citep{Zhao.2017}. In the context of maximum likelihood estimation (MLE), the MSE loss arises from the assumption of i.i.d. standard Gaussian noise. However, this assumption is violated in CT training data since they are often obtained from reconstructions of high-dose or normal-dose measurements. E.g. the choice of the reconstruction algorithm can lead to artifacts, as shown in Figure \ref{fig:ArtifactSamples}. This implies that the reconstruction error for individual pixels is no longer independent. We argue that these dependencies can be better captured by a flow-based model.

Our contributions are twofold: 1) We apply conditional normalizing flows to CT image reconstruction. 2) We propose a hybrid approach, which integrates the physical model by using the filtered back-projection as conditioner.

\subsection{Related Work}

Deep learning methods have been successfully applied to many ill-posed inverse problems such as CT \citep{Arridge.2019}. In particular, end-to-end learned methods have been used. Those methods can be classified in three main groups: post-processing \citep{chen2017convnet_ct}, fully-learned \citep{he2020iradon} and learned iterative algorithms \citep{adler2017}. These end-to-end methods have in common that they learn a parameterized operator ${T_\theta:Y \rightarrow X}$ by optimizing the parameters $\theta$ using training data $\{ (y_i^\delta, x_i^\dagger) \}_{i=1}^N$. Usually, this is done by minimizing the MSE between the ground truth data $x_i^\dagger$ and the reconstruction $T_\theta(y_i^\delta)$ as
\begin{align*}
    \hat{\theta} \in \arg \min_{\theta \in \Theta} \frac{1}{N} \sum_{i=1}^N \Vert T_\theta(y_i^\delta) - x_i^\dagger \Vert^2.
\end{align*} 

Recently, Deep Image Priors were used for CT, achieving promising results in the low-data regime \citep{baguer2020computed}. Similar to our approach is the work of \citep{Adler.11142018}, who employed a Wasserstein GAN to draw samples from the conditional distribution. However, in this approach it is not possible to evaluate the likelihood of the generated samples. \cite{Ardizzone.14.08.2018} have used invertible neural networks to approximate the conditional distribution and to analyze inverse problems. In a subsequent paper this concept was extended to conditional invertible neural networks (cINNs) which yielded good performance in the field of conditional image generation \citep{Ardizzone.742019}. 

\section{Background on Computed Tomography}
\label{sec:Background}

Computed tomography allows for a non-invasive acquisition of the inside of the human body, which makes it one of the most important tools in modern medical imaging \citep{buzug2008computed_tomography}.  CT is a primary example of an inverse problem. The determination of the interior distribution cannot be achieved directly. It has to be inferred from the measured attenuation of X-rays sent through the body.

Current research focuses on reconstruction methods for low-dose CT measurements to reduce the health risk from radiation \citep{shan2019nn_vs_vendor_ct, baguer2020computed}. One strategy to reduce the dose is to measure at fewer angles. This can result in undersampled measurements and therefore in the existence of ambiguous solutions to the inverse problem. Another option is to lower the intensity of the X-ray. This leads to increased Poisson noise on the measurements and adds to the instability of the inversion. In this paper, we test our reconstruction model for the lower intensity case.

\begin{figure}[t]
    \centering
    \scalebox{0.6}{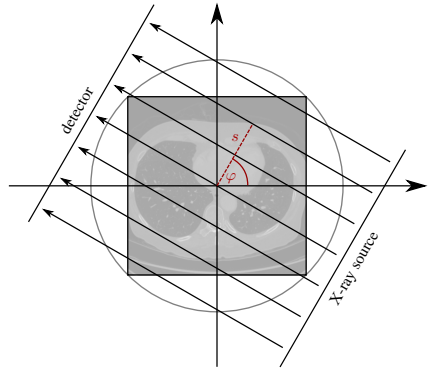}
    \caption{Schematic illustration of a CT scanner with a parallel beam geometry \citep{baguer2020computed}. The scanner is rotated around the patient during the measurement.}
    \label{fig:parallel_beam}
\end{figure}

The basic principle of a CT machine with a parallel beam geometry can be described by the 2D Radon transform {$A: X \rightarrow Y$} \citep{radon1986radon_trafo}:
\begin{equation}
    A x(s,\varphi) = \int_{\mathbb{R}} x\left( s \begin{bmatrix} \cos(\varphi) \\ \sin(\varphi) \end{bmatrix} + t \begin{bmatrix} -\sin(\varphi) \\ \phantom{-}\cos(\varphi) \end{bmatrix} \right) \, \mathrm{d}t.
    \label{eq:radon_trafo}
\end{equation}
It is an integration along a line, which is parameterized by the distance $s \in \mathbb{R}$ and angle $\varphi \in [0,\pi]$ (cf.\ Figure \ref{fig:parallel_beam}). For a fixed pair $(s,\varphi)$ this results in the log ratio of initial and final intensity at the detector for a single X-ray beam (Beer-Lambert's law). The whole measurement, called \textit{sinogram}, is the collection of the transforms for all pairs $(s,\varphi)$. The task in CT is to invert this process to get a reconstruction of the body. The inversion of the Radon transform is an ill-posed problem since the operator is linear and compact \cite{natterer2001math_tomography}. The consequences is an 
instable inverse mapping, which amplifies even small measurement noise.

A common inversion model is the filtered back-projection (FBP)  \citep{Shepp1974FBP}. The reconstruction for position $(i,j)$ is calculated by a convolution over $s$ and an integration along $\varphi$ as
\begin{align*}
    x(i,j) = \int_0^{\pi} y(s, \varphi) \star h(s) \vert_{s = i \cos(\varphi) + j \sin(\varphi)} \, \mathrm{d}\varphi.
\end{align*}
In general, $h$ is chosen as a high-pass filter such as the Ram-Lak filter \cite{Ramachandran2236}. In reality, we can only measure for a finite number of angles and distances. In this discrete setting the FBP only works well for a high number of measurement angles. Otherwise severe streak artifacts can appear in the reconstruction. 

\section{Methods}
\label{sec:methods}

\subsection{Problem Setting}
\label{sec:probSetting}

To estimate conditional densities, data pairs from measurements $y^\delta$ and ground truth images $x^\dagger$ are required. In computed tomography (CT) it is not possible to obtain actual ground truth data, because no picture can be taken of the interior of the human body. Instead of using ground truth images we use reconstructions based on high-dose measurements $y^{\delta_1} = A x^\dagger + \mu_1$, i.e.~$x^{\delta_1} = T_\text{FBP}(y^{\delta_1})$. Because $x^{\delta_1}$ is the output of an reconstruction algorithms, it can contain artifacts and differ from the actual image $x^\dagger$, see Figure \ref{fig:ArtifactSamples} for an example. In the next step we simulate low-dose CT measurements using this reconstruction as $y^{\delta_2} = A x^{\delta_1} + \mu_2$, where $\mathrm{Var}[\mu_2] \geq \mathrm{Var}[\mu_1]$, since low-dose measurements are more prone to measurement noise. The training set then consists of data pairs $\{y^{\delta_2}, x^{\delta_1} \}$. An example of such a dataset is LoDoPaB-CT~\citep{leuschner2019lodopabct}, which we use to benchmark our proposed conditional flow.

\subsection{Normalizing Flow with FBP Conditioning}
From a statistical point-of-view, an inverse problem can also be seen as a generating process $x \sim p(x|y)$ \cite{Dashti.2272013, Arridge.2019}. The task in such a statistical inverse problem is to estimate this conditional distribution. We are using conditional normalizing flows (NF) \citep{Winkler.11292019} to approximate the target density $p(x|y)$. The conditional NF is composed of a series of invertible transformations $F= f_K \circ \dots \circ f_1$. Here, every individual transformation is parameterized by $\theta$ and receives a conditional input $y$: $f_i = f_{\theta_i}(\cdot, y)$. This transformation defines a transport map, which converts the initial density into a simple, easy-to-sample density $p_Z$. This model defines a conditional density $q(x|y,\theta)$ and using the change-of-variables formula the conditional density can be calculated:
\begin{align*}
	q(x|y;\theta) = p_{Z}(F_\theta(x;y)) \left\vert \det\left(\frac{\partial F_\theta(x;y)}{\partial x}\right) \right\vert.
\end{align*}
We denote the Jacobian for one data point $x_i, y_i$ with $J_i = \frac{\partial F_\theta(x_i;y_i)}{\partial x}$. Instead of directly using the measurements $y_i$ as conditional inputs, we propose to employ a reconstruction, e.g. the filtered back-projection $\hat{x}_i = T_{\text{FBP}}(y)$.

Assume a dataset $\{(x_i, y_i)\}_{i=1}^N$ of measurements $y_i$ and reconstructions $x_i$. To approximate the target density $p(x|y)$ a conditional NF $q(x|y, \theta)$ can be trained using the negative log-likelihood as a loss function. Using a standard normal distribution, i.e.\ $p_Z \sim \mathcal{N}(0, I)$, this amounts to minimizing
\begin{align*}
    \mathcal{L}(\theta) = \frac{1}{N} \sum_{i=1}^N \frac{\Vert F_\theta(x_i;T_{\text{FBP}}(y_i))\Vert_2^2}{2}  - \log \left\vert \det J_i \right\vert .
\end{align*}

\subsection{Conditional coupling layers}
We are using the conditional coupling layer from \citep{Ardizzone.742019} to construct a conditional invertible neural network (cINN), which is an extension of the affine coupling layer from \citep{Dinh.27.05.2016}. We propose to integrate the model-driven approach of inverse problems by using the filtered back-projection $\hat{x} = T_\text{FBP}(y)$ as conditional input instead of the raw sinogramm measurements $y$. The input $u=[u_1, u_2]$ to an affine coupling layer is split into two parts and both parts are transformed individually:
\begin{align*}
	v_1 &= u_1 \odot \exp(s_1(u_2, \hat{x})) + t_1(u_2, \hat{x}) \\ \newline
	v_2 &= u_2 \odot \exp(s_2(v_1, \hat{x})) + t_2(v_1, \hat{x}).
\end{align*}
The transformations $s_1, s_2, t_1, t_2$ do not need to be invertible and are modelled as convolutional neural networks (CNNs). The inverse of an affine coupling layer is:
\begin{align*}
	u_1 &= (v_1 - t_1(u_2,\hat{x})) \odot \exp(-s_1(u_2, \hat{x})) \\ \newline
	u_2 &= (v_2 - t_2(v_1,\hat{x})) \odot \exp(-s_2(v_1,\hat{x})).
\end{align*}
The log-determinant of the Jacobian for one affine coupling layer can be calculated as the sum over $s_i$, i.e.\ $\sum_j s_1(u_2, \hat{x})_j + \sum_j s_2(v_1, \hat{x})_j$. A deep invertible network can be built as a sequence of multiple such layers, with a permutation of the dimensions after each layer.

The conditional input $\hat{x}$ is added as an extra input to each transformation in the coupling layer. In practice, an additional conditioning network $H$ is added, so instead of $\hat{x}$ the output $H(\hat{x})$ is used. This conditioning network $H$ is under no architectural constraints and can contain all usual elements (i.e.\ BatchNorm, pooling layer, etc.) of a CNN.

\begin{figure}
\centering
\includegraphics[scale=0.23]{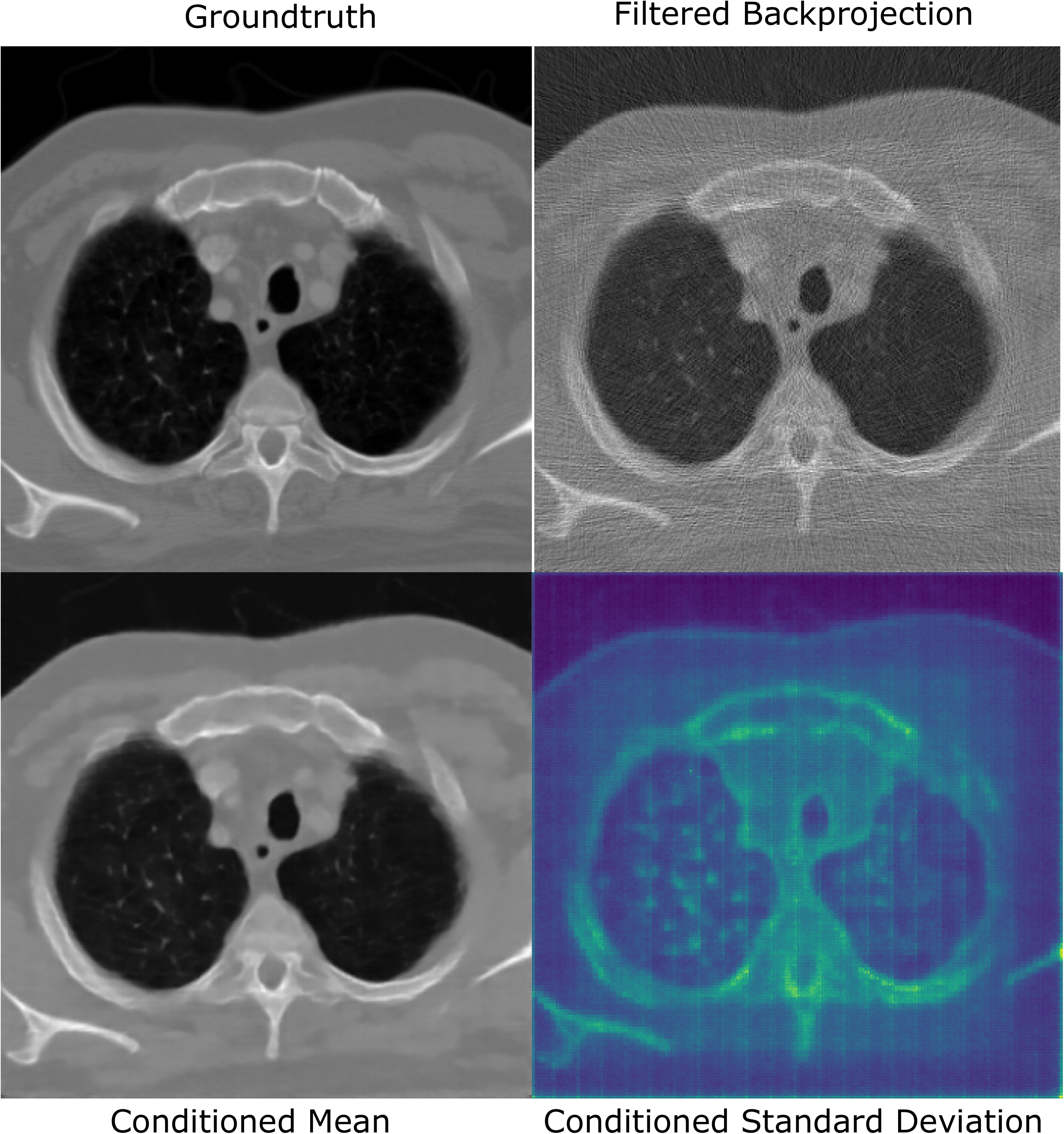}
\caption[ConditionedMean]{Reconstruction and standard deviation of cINN. $1000$ Samples were used for the reconstruction. The top row shows the ground truth image and the corresponding FBP.}
	\label{fig:Reconstruction}
\end{figure}

\section{Results}
\label{sec:results}

Sampling from the model is a two-step process: First, a sample $z$ is drawn from the base density $p_Z$. Second, this sample is transformed by the inverse to obtain an image. By repeatedly sampling $z_j$ for a fixed input $y^\delta$ we thus estimate the conditional mean as
\begin{align*}
    \hat{x} = \mathbb{E}_{z}[F^{-1}(z,T_{\text{FBP}}(y^\delta))] \approx \frac{1}{n} \sum_{j=1}^n F^{-1}(z_j, T_{\text{FBP}}(y^\delta)).
\end{align*}
We evaluate our model on the LoDoPaB-CT dataset \cite{leuschner2019lodopabct}. For this dataset we are in the case of oversampling, so we expect a uni-modal distribution. This enables the choice of the conditional mean as the reconstruction method. If a highly multi-modal distribution is expected, the conditional mean is not the optimal choice. 
To measure the error between reconstruction and ground truth, the PSNR and the SSIM \citep{wang2004ssim} are evaluated. Both are common quality metrics for the evaluation of CT and MRI reconstructions \cite{joemai2017ssim_in_ct, adler2017, zbontar2018dataset_fastmri}. The PSNR is a pixel-wise metric which is defined via the MSE. The SSIM is a structural metric, which compares local patterns of pixels and is not calculated on a per pixel basis.

\subsection{Implementation}
\label{subsec:implementation}
We follow the multi-scale architecture design of RealNVP \citep{Dinh.27.05.2016}. After each block, consisting of 6 coupling layers, downsampling is performed. The downsampling is done using the Haar downsampling from \cite{Ardizzone.742019} and the variant used in \cite{Jacobsen.2018}. The dimensions have to be permuted after each coupling layer. This is done using the invertible 1x1 convolutions from \cite{Kingma.792018}. The model is implemented using the library FrEIA\footnote{https://github.com/VLL-HD/FrEIA}. A conditioning network was used to further extract features from the filtered back-projection. This conditioning network was trained together with the full flow-based model. Details on the implementation can be found in the supplementary material.

\subsection{LoDoPaB-CT Dataset}
\label{subsec:lodopab}
We evaluate our method on the low-dose parallel beam (LoDoPaB) CT dataset \cite{leuschner2019lodopabct}, which contains over \num{40000} two-dimensional CT images and corresponding simulated low photon count measurements.
The ground truth images $x^{\delta_1}$ are human chest CT reconstructions from the LIDC/IDRI database \cite{armato2011lidc_idri}, cropped to $\num{362}\times\num{362}$ pixels.
Projections are computed using parallel beam geometry with \num{1000} angles and \num{513} beams.
Poisson noise is applied to model a low photon count ($\mu_2$ in Section \ref{sec:probSetting}).

\subsection{Evaluation on LoDoPaB-CT}

 We have evaluated our model on the LoDoPaB-CT dataset. First we examined the dependence of PSNR and SSIM on the number of samples for the conditional mean. The results are shown in Figure \ref{fig:SSIMPSNR} (appendix). Both PSNR and SSIM increase with a higher number of samples. This allows for a trade-off between quality of reconstruction and time. For our evaluation we have chosen a conditional mean with 1000 samples. Table \ref{tab:results_lodopab} shows the scores on the test dataset. The comparison includes several classical and deep learning approaches.

In terms of PSNR our model is comparable to other state-of-the-art deep learning approaches, despite not explicitly trained to minimize the MSE between the prediction and the ground truth images. Regarding the SSIM our model outperforms all other approaches. This further underlines our hypothesis that using the more flexible flow objective enable to incorporate structural properties.

\begin{table}[htb]
\centering
\setlength{\tabcolsep}{12pt}
\begin{tabular}{lS[table-format=2.2]S[table-format=1.2]}
\textbf{Model} & \textbf{PSNR} & \textbf{SSIM} \\
\toprule
FBP \scriptsize{\cite{Shepp1974FBP}}& 30.52 & 0.74 \\               
FBP + U-Net \scriptsize{\citep{jin2017cnn_imaging}} & 35.90 & 0.85 \\       
TV Regularization & 32.27 & 0.78 \\                
DIP + TV \scriptsize{\citep{baguer2020computed}} & 34.79 & 0.82 \\          
iRadonMap \scriptsize{\citep{he2020iradon}} & 31.23 & 0.76 \\         
Learned GD \scriptsize{\citep{adler2017iterative_nn}}& 34.67 & 0.82 \\        
Learned PD \scriptsize{\citep{adler2017}} & \cellcolor{black!10}\textbf{36.12} & 0.85 \vspace{0.35em} \\        
\hline
cINN \scriptsize{(Ours)} & 35.68  & \cellcolor{black!10}\textbf{0.88} \\ 

\end{tabular}
\caption{Results on the LoDoPaB-CT test set. The baseline methods were evaluated in \cite{baguer2020computed}.}
\label{tab:results_lodopab}
\end{table}

\section{Conclusion}
\label{sec:conclusion}

We have investigated how flow-based models can be applied as a conditional density estimator for the
reconstruction of low-dose CT images. Using this generative approach, we were able to obtain high-quality reconstructions that outperformed all other deep learning approaches in terms of structural similarity. So far only coupling-based INNs were used, but future work should explore other architectures such as i-ResNets \citep{Behrmann.} for this conditional density estimation task. Furthermore, our hybrid approach that integrates the physical model into the conditioning could enable the use of more advanced reconstruction algorithms. Thus, conditional flows are a promising avenue for statistical model-based inverse problems such as CT reconstruction.

\section{Acknowledgements}
\label{sec:acknowledgements}
Johannes Leuschner and Maximilian Schmidt acknowledge the support by the Deutsche Forschungsgemeinschaft (DFG) within the framework of GRK 2224/1 ``$\pi^3$: Parameter Identification -- Analysis, Algorithms, Applications''.

\bibliography{references.bib}

\section{Appendix}

\subsection{Model Architecture}
The model was trained for \num{15.000} stochastic gradient steps of batchsize \num{10} with the Adam-Optimizer \citep{kingma2014adam} using a weight decay of \num{e-5}. 
The last layer in the subnetworks of each coupling layer is initialized with zero. This initializes the model as a whole with the identity.

A multiscale architecture was used for implementation of the cINN. The model includes 6 resolution levels. After each level a part of the channels is split off and passed on to the output. After each resolution level downsampling is performed. Downsampling was performed using the iRevNet variant \citep{Jacobsen.2018} as well as the Haar Downsampling by  \cite{Ardizzone.742019}. The input size of the CT-images is $1 \times 352 \times 352$. The full cINN was build as follows.

\begin{tabular}{|l|l|}
\hline cINN & Output size \\
\hline  iRevNet-Downsampling & $4 \times 176 \times 176$ \\
\hline level 1 conditional section &  $4 \times 176 \times 176$   \\ 
\hline  iRevNet-Downsampling & $16 \times 88 \times 88$ \\
\hline Split: $8 \times 88 \times 88$ to output & $8 \times 88 \times 88$ \\
\hline level 2 conditional section &  $8 \times 88 \times 88$   \\ 
\hline  iRevNet-Downsampling & $32 \times 44 \times 44$ \\
\hline Split: $16 \times 44 \times 44$ to output & $16 \times 44 \times 44$ \\
\hline level 3 conditional section &  $16 \times 44 \times 44$   \\ 
\hline  iRevNet-Downsampling & $64 \times 22 \times 22$ \\
\hline Split: $32 \times 22 \times 22$ to output & $32 \times 22 \times 22$ \\
\hline level 4 conditional section &  $32 \times 22 \times 22$   \\ 
\hline  Haar-Downsampling & $128 \times 11 \times 11$ \\
\hline Split: $96 \times 11 \times 11$ to output & $32 \times 11 \times 11$ \\
\hline level 5 conditional section &  $32 \times 11 \times 11$   \\ 
\hline Split: $28 \times 11 \times 11$ to output & $4 \times 11 \times 11$ \\
\hline level 6 Dense-conditional section & $484$ \\
 \hline 
\end{tabular}{}

A conditioning network was used to extract features from the conditional input. Similar to the cINN, this network consists of 6 resolution levels. The output from the resolution level of the conditioning network is used as the conditioning input for the respective resolution level in the cINN. If not specified otherwise, a kernel size of $k=3$ is used. In addition, batch normalization (BN) is applied.

\begin{tabular}{|l|}
\hline
Conv2d: $1 \rightarrow 3$, stride=2 + LeakyReLU    \\ 
\hline Conv2d: $32 \rightarrow 64$ + LeakyReLU \\
\hline Conv2d: $64 \rightarrow 128$ + BN + LeakyReLU \\
\hline Conv2d: $128 \rightarrow 64$ + BN + LeakyReLU \\
\hline Conv2d: $64 \rightarrow 32$ + BN + LeakyReLU \\
\hline Conv2d: $32 \rightarrow 4$ + BN ($\rightarrow$ level $1$) \\
 \hline
\end{tabular}{}

\begin{tabular}{|l|}
\hline
LeakyRelu \\
\hline Conv2d: $4 \rightarrow 32$, stride=2 + BN + LeakyReLU    \\ 
\hline Conv2d: $32 \rightarrow 32$ (k=1) + LeakyReLU \\
\hline Conv2d: $32 \rightarrow 32$ + BN + LeakyReLU \\
\hline Conv2d: $32 \rightarrow 8$ + BN ($\rightarrow$ level $2$)\\
 \hline
\end{tabular}{}

\begin{tabular}{|l|}
\hline
LeakyRelu \\
\hline Conv2d: $8 \rightarrow 32 $ (k=1) +  LeakyReLU    \\ 
\hline Conv2d: $32 \rightarrow 64$, stride = 2+ LeakyReLU \\
\hline Conv2d: $64 \rightarrow 16$ + BN ($\rightarrow$ level $3$)  \\
 \hline
\end{tabular}{}

\begin{tabular}{|l|}
\hline
LeakyRelu \\
\hline Conv2d: $16 \rightarrow 64 $ (k=1) +  LeakyReLU    \\ 
\hline Conv2d: $64 \rightarrow 64$, stride = 2+ LeakyReLU \\
\hline Conv2d: $64 \rightarrow 32$ + BN ($\rightarrow$ level $4$) \\
 \hline
\end{tabular}{}

\begin{tabular}{|l|}
\hline
LeakyRelu \\
\hline Conv2d: $32\rightarrow 96 $ (k=1) +  LeakyReLU    \\ 
\hline Conv2d: $96 \rightarrow 128$, stride = 2+ LeakyReLU \\
\hline Conv2d: $128 \rightarrow 32$ (k=1) + BN  ($\rightarrow$ level $5$)\\
 \hline
\end{tabular}{}

\begin{tabular}{|l|}
\hline
LeakyRelu \\
\hline Conv2d: $32\rightarrow 64 $, stride=2 +  LeakyReLU    \\ 
\hline Conv2d: $64 \rightarrow 256$, stride = 2+ LeakyReLU \\
\hline Average Pooling + Flatten + BN ($\rightarrow$ level $6$) \\
 \hline 
\end{tabular}{}

To implement the subnetworks in the coupling layers a CNN variant and a fully connected variant were used. The input channels are denoted by $c_{in}$ and the output channels by $c_{out}$.

\begin{tabular}{|l|}
\hline CNN-subnetwork (k=1) or (k=3) \\
\hline Conv2d: $c_{in}\rightarrow 64 $, +  LeakyReLU    \\ 
\hline Conv2d: $64 \rightarrow 92$ + LeakyReLU \\
\hline Conv2d: $92 \rightarrow c_{out}$ \\
 \hline 
\end{tabular}{}

\begin{tabular}{|l|}
\hline Dense-subnetwork  \\
\hline Dense: $c_{in}\rightarrow 512 $, +  LeakyReLU    \\ 
\hline Dense: $512 \rightarrow 512$ + LeakyReLU \\
\hline Dense: $512 \rightarrow c_{out}$ \\
 \hline 
\end{tabular}{}

Using this two variants of subnetworks the conditional sections are implemented as follows.

\begin{tabular}{|l|l|}
\hline
\multicolumn{2}{|l|}{conditional section} \\ \hline
 Coupling (CNN-subnet k=1) & \multirow{4}{*}{3x}          \\ \cline{1-1}
 Glow $1 \times 1$ convolution &   \\ \cline{1-1}
 Coupling (CNN-subnet k=3)  &   \\ \cline{1-1}
 Glow $1 \times 1$ convolution & \\ \hline
\end{tabular}

\begin{tabular}{|l|l|}
\hline
\multicolumn{2}{|l|}{dense conditional section} \\ \hline
   Random permutation & \multirow{2}{*}{4x}             \\ \cline{1-1}
   Dense-subnetwork &   \\ \hline
\end{tabular}

After each downsampling a small unconditioned subnetwork is used: 

\begin{tabular}{|l|}
\hline CNN-subnetwork (without conditional input) \\
\hline Conv2d: $c_{in}\rightarrow 64$ (k=1), +  LeakyReLU    \\ 
\hline Conv2d: $64 \rightarrow 64$ (k=1) + LeakyReLU \\
\hline Conv2d: $64 \rightarrow c_{out}$ (k=1)\\
 \hline 
\end{tabular}{}

The downsampling section is built as follows: 

\begin{tabular}{|l|}
\hline Downsample section (Haar or iRevNet) \\
\hline Haar or iRevNet downsampling   \\ 
\hline Glow $1 \times 1$ convolution\\
\hline Coupling (unconditional CNN-subnetwork)\\
\hline Glow $1 \times 1$ convolution\\
\hline Coupling (unconditional CNN-subnetwork)\\
 \hline 
\end{tabular}{}

\subsection{Evaluation of the Conditional Mean}
\begin{figure}[h]
\centering
\includegraphics[width=\columnwidth]{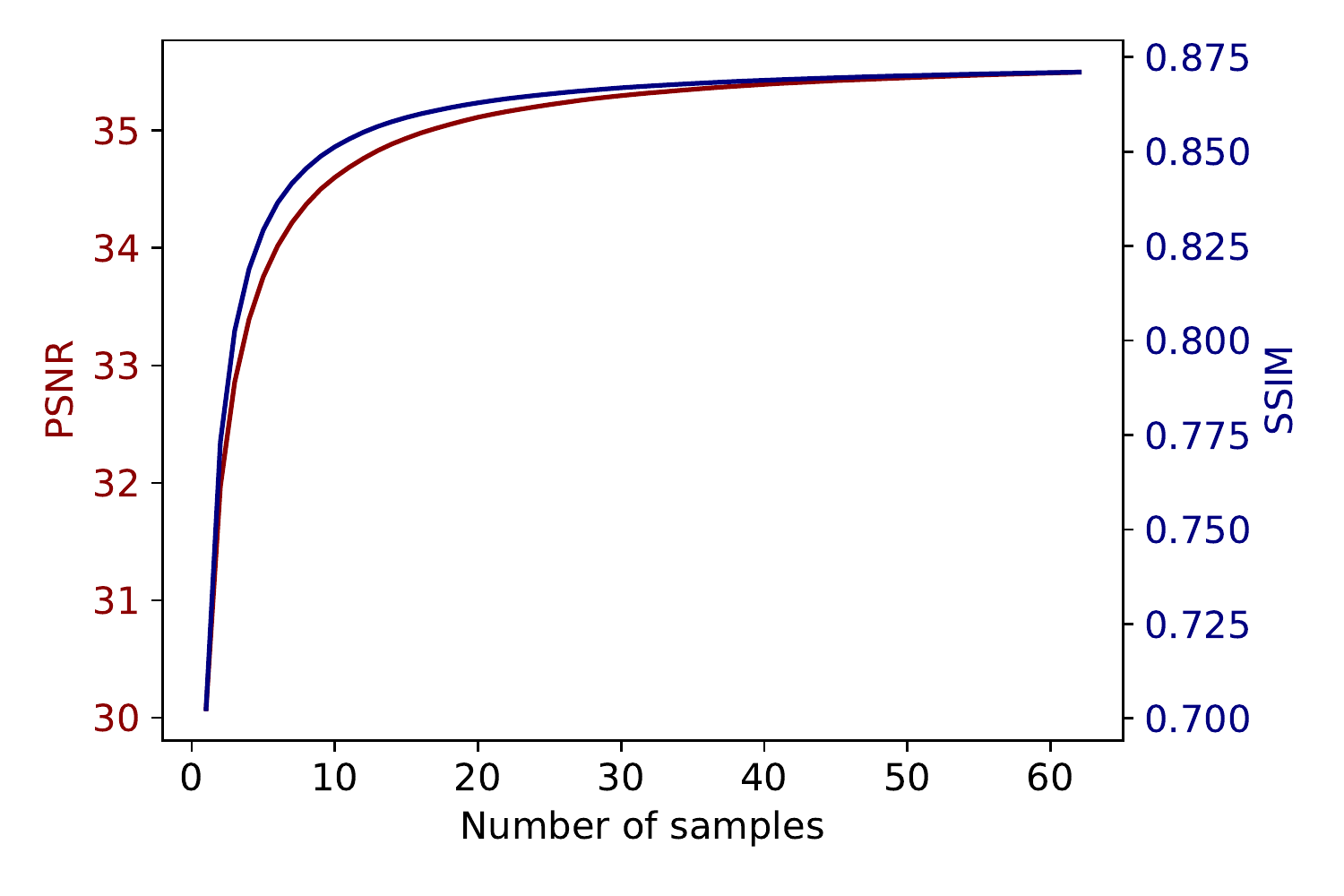}
\caption[SSIM/PSNR by number of samples]{
PSNR and SSIM for the validation set of the LoDoPaB-CT dataset. The PSNR is colored in red and the SSIM is colored in blue.}
	\label{fig:SSIMPSNR}
\end{figure}

We have used the conditional mean as a reconstruction for the CT image. Figure \ref{fig:SSIMPSNR} shows the performance in relation to the number of samples used. 

\newpage

\onecolumn
\subsection{Additional Examples}

\begin{figure}[h]
\centering
\includegraphics[scale=0.21]{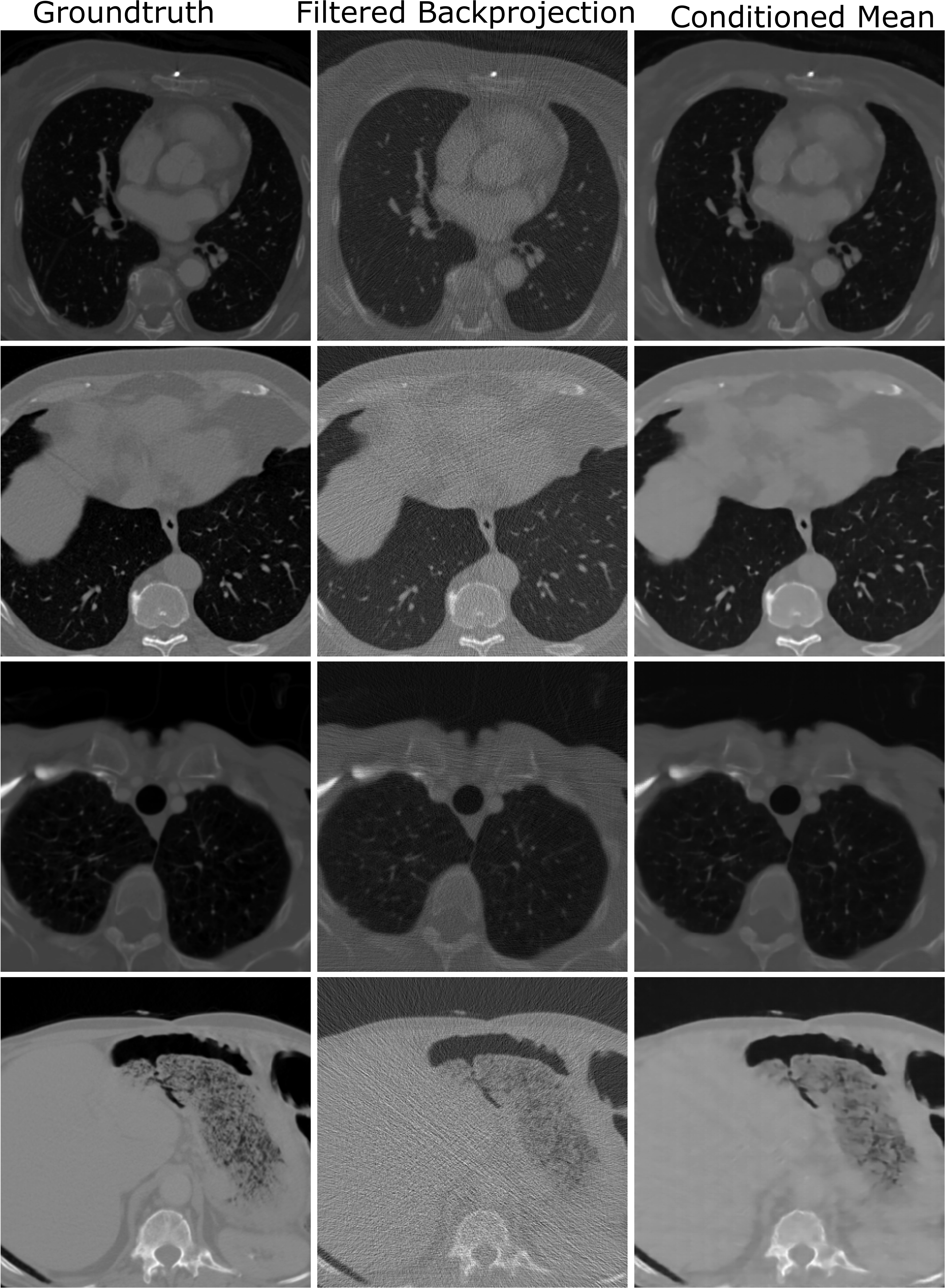}
	\label{fig:Reconstruction2}
\end{figure}

\end{document}